\documentclass[aps,prl,reprint,floatfix,longbibliography]{revtex4-1}
\usepackage{graphicx}
\usepackage{amsmath}
\usepackage[english]{babel}

\newcommand*{\pr}[1]{\mathcal{#1}}
\newcommand*{\refneq}[1]{(\ref{#1})}
\newcommand*{\refneqs}[1]{(\ref{#1})}
\newcommand*{\refeq}[1]{Eq.\ (\ref{#1})}

\newcommand*{\refsup}{\cite{supplement}}

\begin{document}


\title{Phase transitions in definite total spin states of two-component Fermi gases}


\author{Vladimir A. Yurovsky}
\affiliation{School of Chemistry, Tel Aviv University, 6997801 Tel Aviv, Israel}

\date{\today}

\begin{abstract}
Second-order phase transitions have no latent heat and are characterized by a change in symmetry. In addition to the conventional symmetric and anti-symmetric states under permutations of bosons and fermions, mathematical group-representation theory allows for non-Abelian permutation symmetry. Such symmetry can be hidden in states with defined total spins of spinor gases, which can be formed in optical cavities. The present work shows that the symmetry reveals itself in spin-independent or coordinate-independent properties of these gases, namely as non-Abelian entropy in thermodynamic properties. In weakly interacting Fermi gases, two phases appear associated with fermionic and non-Abelian symmetry under permutations of particle states, respectively. The second-order transitions between the phases are characterized by discontinuities in specific heat. Unlike other phase transitions, the present ones are not caused by interactions and can appear even in ideal gases. Similar effects in Bose gases and strong interactions are discussed. 
\end{abstract}


\maketitle
A distinctive feature of phase transitions is analytic discontinuities or singularities in the thermodynamic functions 
\cite{pathria}. The transitions, analyzed here, are related to the permutation symmetry. 
According to the Pauli exclusion principle, the many-body wavefunction can be either symmetric of anti-symmetric over particle permutations \cite{landau}. The particles can be either elementary --- like electrons or photons --- or composite --- as atoms and molecules.

The symmetric and anti-symmetric wavefunctions belong to one-dimensional irreducible representations (irreps) of the symmetric (or permutation) group \cite{hamermesh}. However, group theory allows for the multidimensional, non-Abelian irreps of this group. They can be illustrated by many-body spin wavefunctions of electrons. A two-electron system with the total spin projection $0$ has two states. In the first one, the first and the second electrons are in the spin up and spin down states, respectively, and vice versa in the second state. These two states can be symmetrized or anti-symmetrized, giving the triplet and singlet states, respectively.

In the case of three electrons with the total spin projection $1/2$, each of them can be in the spin down state. This provides three non-symmetric states. Symmetrization over permutations provides a one-dimensional irrep. However, the anti-symmetric state does not exist, since two electrons are in the same spin up state. Then two three-body wavefunctions, which are orthogonal to the symmetric wavefunction, form a two-dimensional irrep.

Non-Abelian permutation symmetry has been considered in early years of quantum mechanics by Wigner \cite{wigner1927}, Heitler \cite{heitler1927}, and Dirac \cite{dirac1929}, before the Pauli exclusion principle was discovered. Particles with such symmetry, called ``intermedions'' were considered later and there are strong arguments that the total wavefunction cannot belong to a non-Abelian irrep \cite{kaplan2013}. Nevertheless, if the spin and spatial degrees of freedom are separable, the total wavefunction, satisfying the Pauli principle, can be represented as a sum of products of spin and spatial wavefunctions with non-Abelian permutation symmetry. (Such wavefunctions are used in spin-free quantum chemistry \cite{kaplan,pauncz_symmetric}, one-dimensional systems \cite{yang1967,sutherland1968} and molecular relaxation \cite{brechet2016}.) Then spin-independent or coordinate-independent properties of such systems will be the same as ones of hypothetical intermedions. The present work analyses unusual thermodynamic properties arising from non-Abelian permutation symmetry. 

A wavefunction can be symmetric or antisymmetric for any number of particles $N$. In contrast, the non-Abelian irrep matrices are specific for each $N$. Then the non-Abelian case can be described in canonical and microcanonical ensembles, but not in a grand-canonical one.  
In a microcanonical ensemble \cite{pathria}, the macrostate of the gas is determined by $N$, the total energy $E$, the external potential or the volume where the particles are contained, and, in the present case, by the many-body spin $S$. According to the postulate of equal \textit{a priory} probabilities \cite{pathria}, the system is equally likely to be in any microstate consistent with given macrostate. The microstates are eigenstates of the many-body Hamiltonian. (An alternative derivation \refsup    is based on the Berry conjecture \cite{berry1977} rather than on the postulate of equal \textit{a priory} probabilities.) 

Randomization of phases, due to either Hamiltonian chaos (as expressed by the Berry's conjecture \cite{berry1977,srednicki1994}) or interactions with the environment, allows us to perform any unitary transformation of the microstates \cite{pathria}, namely, to eigenstates of non-interacting particles. For a gas of spin-$1/2$ fermions they are eigenstates of the Hamiltonian
\begin{equation}
\hat{H}=\hat{H}_{\mathrm{spin}}+\hat{H}_{\mathrm{spat}}, 
\label{Hamiltonian}
\end{equation}  
where $\hat{H}_{\mathrm{spin}}$ is independent of the particle coordinates and $\hat{H}_{\mathrm{spat}}$ is spin-independent. Since the Hamiltonian \refneqs{Hamiltonian} contains no terms that depend on both spins and coordinates, its eigenstates have the defined total spin $S$ and can be represented as \refsup
\begin{equation}
\tilde{\Psi}_{\hat{r}\{\varepsilon\}}^{(S)}=f_{S}^{-1/2}(N)\sum_{t}\tilde{\Phi}_{t\hat{r}\{\varepsilon\}}^{(S)}\Xi_{t}^{(S)}.
\label{tilPsi}
\end{equation}
Here the spatial $\tilde{\Phi}_{t\hat{r}\{\varepsilon\}}^{(S)}$ and spin $\Xi_{t}^{(S)}$ wavefunctions belong to conjugate irreps of the symmetric group. The irreps are associated with  the Young diagram $[2^{N/2-S}, 1^{2S}]$, which is  pictured as $N/2-S$ rows with $2$ boxes and $2S$ rows with 1 box [see, e.g, Figs. \ref{Fig_pop2} (a) and (b)]. The Young diagram is unambiguously determined by the total spin $S$ and the irreps have the dimension $f_{S}(N)$ \refsup.

The functions within irreps are labeled by the standard Young tableaux $t$ --- the Young diagram $[2^{N/2-S}, 1^{2S}]$ filled by the numbers $1\ldots N$ which increase down each column and right each row \refsup.  The microstates are specified by the set of single-body energies $\{\varepsilon\}\equiv\{\varepsilon_1\ldots \varepsilon_N\}$ and the Weyl tableau $\hat{r}$ \cite{harshman2016a}. The latter is a two-column Young diagram $[2^{N/2-S}, 1^{2S}]$ filled by $\varepsilon_j$ such that they increase down each column but may be equal or increase right each row [see Figs. \ref{Fig_pop2} (a) and (b)]. Then in the case of spin-$1/2$ fermions the set $\{\varepsilon\}$ can contain no more than double degeneracies.
As proved in \refsup, the tableau $\hat{r}$ can take $f_S(q_1)$ values, where $q_1$ is the number of non-degenerate energies in the set $\{\varepsilon\}$. Then $f_S(q_1)$ can be considered as a statistical weight of the many-body state. Since the energies have to increase down the columns, the degenerate energies have to be placed in different columns, and the number of pairs of equal $\varepsilon_j$, $q_2=(N-q_1)/2$, can not exceed the shorter column length $N/2-S$.

The eigenstates \refneq{tilPsi} with a defined total spin form a set of degenerate states with collective spin wavefunctions $\Xi_{t}^{(S)}$ and undefined spin projections of individual particles. The Hamiltonian \refneq{Hamiltonian} has also a set of degenerate eigenstates with the same energy, but with defined individual spin projections and an undefined total spin. Given the total spin projection $S_z$ (sum of individual spin projections), these sets can be connected by a unitary transformation.

Spin-independent interactions between particles split energies of the states with different total spins, making the set with defined individual spins inapplicable \cite{heitler1927}, but this effect is small for weakly-interacting gases. A particular case of the states with defined total spins is the collective Dicke states \cite{dicke1954} of two-level particles, coupled by electromagnetic field in a cavity. A two-dimensional cavity leads to spin-dependent spatially-homogeneous interactions of the form \cite{sela2016} 
$\hat{H}_{\mathrm{spin}}=I \hat{S}_+  \hat{S}_-$, where $\hat{S}_+$ and $\hat{S}_-$ are the total spin raising and lowering operators. Such interaction, realized in recent experiments \cite{landig2016}, lead to the energy shift 
$E_{S S_z}=I[S_z(S_z-1)-S(S+1)]$, providing substantial splitting of the states with different total spins \refsup.

The protocol, proposed in \cite{yurovsky2014}, starts from the spin-polarized state with $S=S_z=N/2$. A time-dependent potential, which changes the spin states of particles, but, being coordinate independent, conserves the total spin, can transfer the population to the state with $S=N/2$, $S_z=N/2-1$. Later a potential, which does not change the spin states of particles, can, being dependent on coordinates and spins, transfer the population to the state with $S=N/2-1$, $S_z=N/2-1$. A sequence of such pulses with proper time-dependencies can populate the state with any total spin. The population will not be transferred back to higher $S$ and $S_z$, since the energy spectrum $E_{S S_z}$ is not equidistant and, therefore,  
$E_{S S_z}-E_{S S_z-1}\neq E_{S S_z+1}-E_{S S_z}$ and $E_{S S_z}-E_{S-1 S_z}\neq E_{S+1 S_z}-E_{S S_z}$.
\begin{figure}
\includegraphics[width=3.4in]{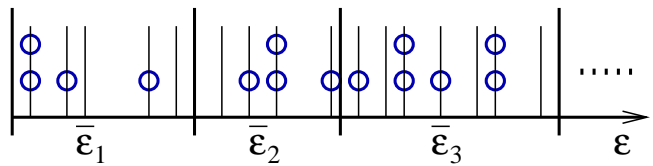}
 \caption{Cells with average energies $\bar{\varepsilon}_i$ in a single-body energy spectrum. The circles denote the level occupation. \label{Fig_ecell}}
\end{figure}

Following the Gentile's version \cite{gentile1940} of the general microcanonical approach, let us divide the single-body energy spectrum into cells (see Fig. \ref{Fig_ecell}) containing $g_i$ energy levels with the average energy $\bar{\varepsilon}_i$. Let $q^{(i)}_0$, $q^{(i)}_1$, and $q^{(i)}_2$ levels be, respectively, non-, single-, and double-occupied in the $i$th cell. Given these occupations, the levels in the cell can be distributed in $g_i!/(q^{(i)}_0!q^{(i)}_1!q^{(i)}_2!)$ distinct ways \cite{gentile1940}. Then the number of distinct microstates associated with the sets $q^{(i)}_l$ is $f_S(q_1)\prod_ig_i!/(q^{(i)}_0!q^{(i)}_1!q^{(i)}_2!)$. 
The system configuration corresponds to the most-probable values of $q^{(i)}_l$ \refsup. They maximize the number of microstates, or its logarithm --- entropy
\begin{equation}
H=\sum_i\left[ g_i \ln g_i-\sum_{l=0}^2 q^{(i)}_l\ln q^{(i)}_l\right] +\ln f_S(q_1) .
\label{entropy}
\end{equation}
Here the Stirling approximation is used. The number of non-degenerate energies $\varepsilon_j$ in the set $\{\varepsilon\}$ is equal to the total number of single-occupied levels $q_1=\sum_{i}q^{(i)}_1$.
The sum in \refeq{entropy} gives the entropy of the Gentile gas \cite{gentile1940}. The present results follow from the last term, which will referred to as non-Abelian entropy, since it vanishes when $f_S=1$.
\begin{figure}
\includegraphics[width=3.4in]{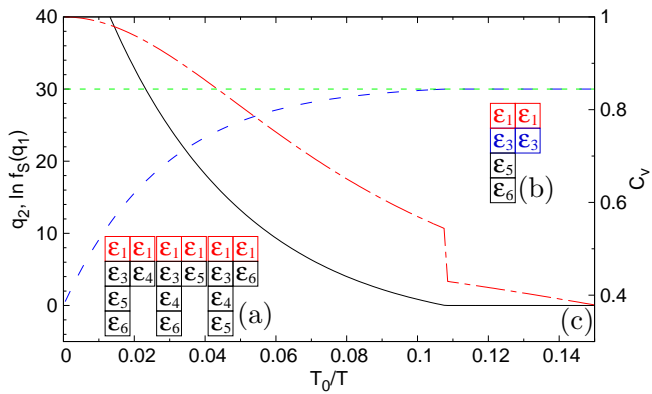}
 \caption{(a) Three allowed Weyl tableaux for $\varepsilon_1=\varepsilon_2<\varepsilon_3<\varepsilon_4<\varepsilon_5<\varepsilon_6$ corresponding to the unsaturated phase. The black cells form Young tableaux corresponding to a non-Abelian irrep. (b) A Weyl tableau for $\varepsilon_1=\varepsilon_2<\varepsilon_3=\varepsilon_4<\varepsilon_5<\varepsilon_6$  corresponding to the saturated phase. The black cells form a one-column Young tableau corresponding to an antisymmetric irrep. (c) The total number of double-occupied levels $q_2$ (blue long dash), the maximal allowed value of  $q_2$ (green horizontal short dash),  non-Abelian entropy $\ln f_S(q_1)$ (black solid line), and specific heat (per atom) $C_v$ (red dot-dashed line) at the temperature $T$ for $N=10^2$ two-dimensional particles in a flat potential with the total spin $S=20$. The temperature scale $T_0$ is given by \refeq{Tscale}.  \label{Fig_pop2}}
\end{figure}
\begin{figure}
\includegraphics[width=3.4in]{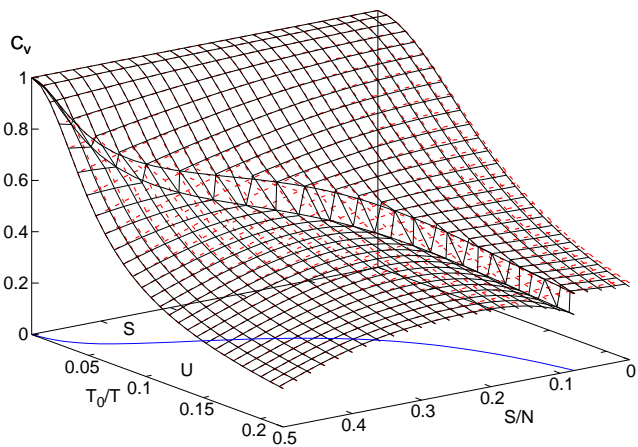}
 \caption{Specific heat (per atom) at the temperature $T$ for the state with the defined many-body spin $S$ (black solid lines) and the state with defined individual spins and the total spin projection $S_z=S$ (red dashed lines) of $N=10^2$ two-dimensional particles in a flat potential. The blue line shows the boundary between the saturated (S) and unsaturated (U) phases. The temperature scale $T_0$ is given by \refeq{Tscale}. \label{Fig_surf}}
\end{figure}

A permutation of single-body energies in the set $\{\varepsilon\}$ transforms \cite{dirac1929,kaplan} the wavefunction \refneq{tilPsi} to a linear combination of $\tilde{\Psi}_{\hat{r}\{\varepsilon\}}^{(S)}$ with different $\hat{r}$. The Weyl tableaux $\hat{r}$ are unambiguously related to the Young tableaux of the shape $[2^{N/2-S-q_2},1^{2S}]$ obtained by the crossing out of the $q_2$ degenerate pairs of $\varepsilon_j$ from the Weyl tableaux with $N/2-S$ two-box rows \refsup. Then the wavefunctions $\tilde{\Psi}_{\hat{r}\{\varepsilon\}}^{(S)}$ form an irrep, associated with the Young diagram $[2^{N/2-S-q_2},1^{2S}]$, of the group $S_{q_1}$ of permutations of non-degenerate $\varepsilon_j$. 
In the saturated phase, $q_2= N/2-S$, the diagram has one column [see Fig. \ref{Fig_pop2}(b)], the irrep is Abelian, and the many-body state has the statistical weight $f_S(2S)=1$. The unsaturated phase ($q_2< N/2-S$) corresponds to the non-Abelian irreps [see Fig. \ref{Fig_pop2}(a)]. At high temperatures, when the number of double-occupied levels $q_2$ is small, the system is in the unsaturated phase. On the temperature decrease, $q_2$ increases, while the statistical weight $f_S(q_1)$ decreases [see Fig. \ref{Fig_pop2}(c)]. At the critical temperature  $q_2$ reaches the maximal allowed value $N/2-S$, the system transforms to the saturated phase, and  $f_S(q_1)$ has a corner.
This leads to discontinuity of the specific heat (per atom)  $C_v=(\partial E/\partial T)_V/N$ (see Figs. \ref{Fig_pop2}(c) and \ref{Fig_surf} and \refsup). The transition is characterized by the non-Abelian entropy $\ln f_S(q_1)$, which ranges between zero in the saturated phase and nonzero in the unsaturated one. However, $\ln f_S(q_1)$ is not a local order parameter. Rather, it is a topological characteristic of the collective state.

The conventional state with defined individual spins is a mixture of two gases containing $N/2+S_z$ and $N/2-S_z$ particles, respectively, with Fermi-Dirac distributions. It is a superposition of all states with defined total spins $S \le  S_z$. As the statistical weight $f_S(N)$ attains its maximum at $S=\sqrt{N+2}/2$, the state with $S=S_z$ dominates in this superposition, unless $S_z\lesssim \sqrt{N}$. However, thermodynamic properties of each $S$-component in this superposition are determined by the maximum of the mixture entropy, which is different from \refeq{entropy}. Then none of the $S$-components is in its thermal equilibrium. As a result, thermodynamic properties of the mixture and of the non-Abelian state with $S=S_z$ are different, and the mixture does not demonstrate the phase transition (see Fig. \ref{Fig_surf} and \refsup).

\begin{figure}
\includegraphics[width=3.4in]{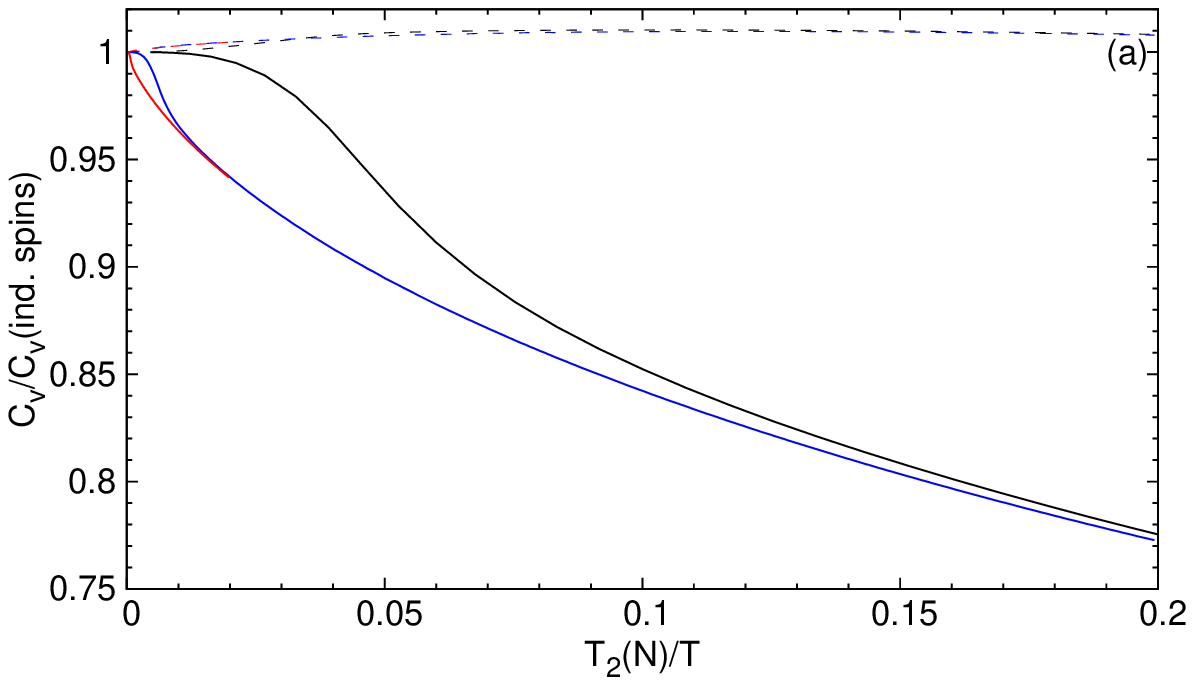}
\includegraphics[width=1.6in]{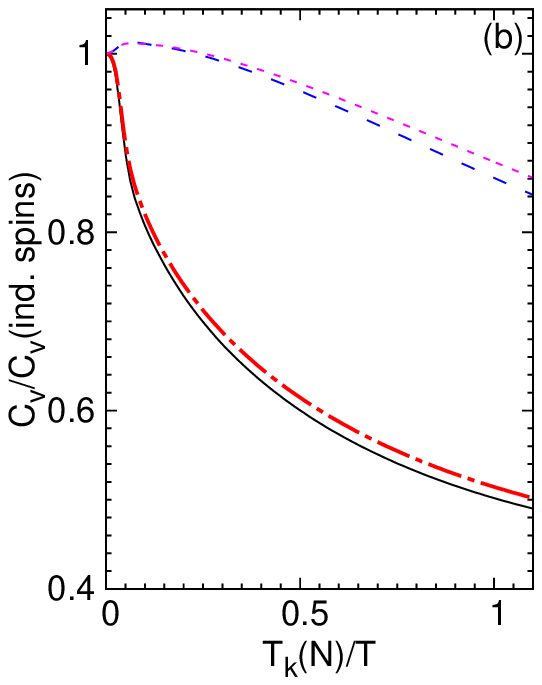}
\includegraphics[width=1.6in]{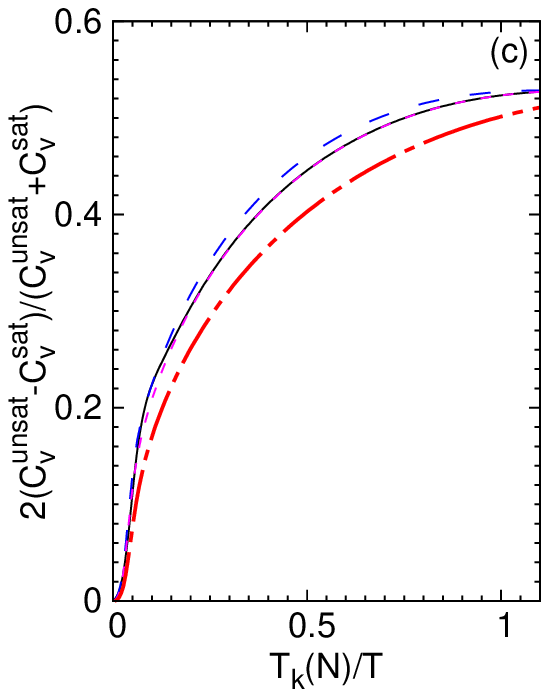}
 \caption{(a) The ratio of the specific heat on the phase boundary to the one for the gas with defined individual spins in a three-dimensional (3D) harmonic trap for the saturated (solid lines) and unsaturated (dashed lines) phases with $N=10^2$ (black), $N=10^3$ (blue) and $N=10^4$ (red) particles. The temperature scale $T_k(N)$ is defined by \refeq{Tscale}. (b) The same ratios for the 3D gas in a flat potential, the black solid and blue long-dashed lines correspond to the saturated and unsaturated phases, respectively. The ratios for a two-dimensional (2D) harmonic trapping are plotted by the red dot-dashed and magenta short-dashed lines, respectively. All plots are for $N=10^2$. (c) The relative change in the specific heat at the phase boundary for $N=10^2$ particles in flat potentials (black solid and blue long-dashed lines for the 2D and 3D cases, respectively) and in  harmonic traps (red dot-dashed and magenta short-dashed lines, respectively).  \label{Fig_crit}}
\end{figure}

The present phase transition has no latent heat since the energy, as well as entropy and pressure, is continuous \refsup. It is therefore a second-order phase transition, like the well-known superconducting one in the absence of magnetic fields. However, the latter is a result of interactions between particles, while the present phase transition can take place in an ideal gas. In this sense, it is similar to the Bose-Einstein condensation phase transition, where the specific heat is discontinuous in the special case of a gas in a 3D harmonic trap \cite{pitaevskii,pethick}. In contrast, the present phase transition takes place in trapped and free gases of any dimension (see Fig. \ref{Fig_surf} and \refsup). Figures \ref{Fig_crit} (a) and (b) show the specific heat at the phase boundary, which is discontinuous and different from the one for defined individual spins. Being plotted as a function of the scaled temperature $T/T_k(N)$, it demonstrates small variation when the trapping and dimensionality are changed (see Figs.  \ref{Fig_crit} (b) and (c)). Here the temperature scale is
\begin{equation}
 T_k(N)=\nu_k^{-1/(k+1)}N^{-k/(k+1)}
 \label{Tscale}
\end{equation} 
and $k$ is the parameter in the energy-density of single-body levels $\nu(\varepsilon)=\nu_k \varepsilon^k$ (for the 2D and 3D gases in flat potentials $k=0$ and $1/2$, respectively, while $k=1$ and $2$ for the 2D and 3D harmonic trapping, respectively, \refsup). The plots for different numbers of particles converge on the decrease of the scaled temperature [see Fig.  \ref{Fig_crit} (a)]. The temperature scale is related to the Fermi energy defined by the equation
$\int_0^{\varepsilon_F} \nu_k \varepsilon^k d\varepsilon=N$ as $\varepsilon_F=[(k+1)N/\nu_k]^{1/(k+1)}$. Then the average energy density $\varepsilon_F/N$ is, up to a factor, the temperature scale \refneq{Tscale}. Figure \ref{Fig_crit} (c) shows that the relative change of the specific heat at the phase boundary approaches $0.5$ at $T<T_k(N)$ for any trapping and dimensionality. Except of the case of a free 2D gas, the temperature scale decreases with increase of $N$. Then the more particles are in the gas, the lower the temperature required in order to observe the phase transition. Even in a free 2D gas, the required temperature decreases in the thermodynamic limit, when $N \to \infty$ with the fixed density $N/V_{2D}$, since $\nu_0\propto V_{2D}$ tends to infinity \refsup and, therefore, $T_0(N)\to 0$. In this sense, the phase transition is a mesoscopic effect (see the discussion in the end of \refsup).

In Gentile's intermediate statistics \cite{gentile1940}, each single-body state can be occupied by a limited number of particles. If this limit is two, Gentile's statistics leads to \refeq{entropy} with $f_S\equiv 1$ and $S=0$, when the two columns of the Young diagram have equal length. For $S=0$, as demonstrated above, the transition temperature tends to zero and the gas is in the unsaturated phase at finite temperatures. Then the phase transition, considered here, cannot appear in Gentile's statistics. Another reason is that the condition  $f_S\equiv 1$ eliminates the non-Abelian entropy and any connection between occupations of single-body states. The non-Abelian entropy depends on the total number of single-occupied states and is not an extensive nor an intensive property, being related to the collective state of the gas.

Zero-range two-body interactions in cold spin-$1/2$ Fermi gases are spin-independent, since collisions of atoms in the same
spin state are forbidden by the Pauli principle. The interactions become spin-dependent and spin and spatial degrees of freedom become inseparable due to inapplicability of the zero-range approximation when the de Broglie wavelength becomes comparable to the effective interaction radius $r_{\text{eff}}$ \refsup. Then the atom energy is  restricted by $\sim 40$mK for ${}^6$Li atoms (the limiting energy is inversely proportional to the atom mass). Under the same condition, the gas can be considered as weakly-interacting and the formation of dimers or Cooper pairs for repulsive or attractive interactions, respectively, can be neglected \refsup, since the elastic scattering length is $|a_S|\approx r_{\text{eff}}$ for non-resonant interactions.

However, the spin and spatial degrees of freedom can be separated for interactions of arbitrary strength while they are spin-independent, and the gas can be kept in a state with the defined many-body spin. For example, in the case of cold atoms, Feshbach resonances \cite{pitaevskii,pethick,chin2010} can provide large $a_S$ for zero-range interactions, leading to non-negligible formation of dimers or Cooper pairs. Since they are symmetric over permutations of forming-particle's coordinates, the number of dimers and Cooper pairs will be restricted by $N/2-S$. This can lead to phase transitions in strongly-interacting gases too, although particles do not occupy single-body states.

In high-spin Fermi gases, similar phase transitions can appear when the interactions are spin-independent, as in $SU(n)$ gases \cite{honerkamp2004,gorshkov2010,cazalilla2009,zhang2014,pagano2014,scazza2014}. If the spatial state of such gas is associated with a Young diagram with non-equal column lengths, a phase transition can be expected when the number of levels occupied by $l$ particles approaches the $l$th column length.

Bose-gases with spin-independent interactions allow for the separation of spin and spatial degrees of freedom, and their states can be associated with Young diagrams too. Such states of spin-$1/2$ bosons were analyzed \cite{kuklov2002,ashhab2003} using $SU(2)$ symmetry (irreps of $SU(2)$ and symmetric groups are closely related, having common basic functions). In the ground state, all particles occupy two lowest levels \cite{kuklov2002,ashhab2003}. Non-Abelian entropy can lead to a phase transition when the occupation of the lowest level approaches the first row length $N/2+S$. For high-spin bosons, phase transitions can be expected when the occupation  of $n$ th excited level approaches the length of $n+1$ th row. A certain analogy can be drawn to the phase transitions in coupled tubes controlled by the tube filling factors \cite{safavi2014}.

States with non-Abelian symmetry can find applications in quantum metrology, computing and information processing, like non-Abelian anyons related to representations of the braid group \cite{stern2013,albrecht2016}.  Thermodynamical properties of an ideal gas of non-Abelian anyons studied in \cite{mancarella2013} do not demonstrate phase transitions.

In conclusion, eigenstates of two-component Fermi gases have defined many-body spins and can be associated with multidimensional, non-Abelian irreps of the symmetric group. An additional energy-degeneracy of the eigenstates modifies the system entropy, leading to second-order phase transitions in the case of weak interactions.

\begin{acknowledgments}
This research was supported in part by a grant from the United States-Israel Binational Science Foundation (BSF) and the United States National Science Foundation (NSF). The author gratefully acknowledge useful conversations with N. Davidson, I. G. Kaplan, and M. Olshanii.
\end{acknowledgments}

\clearpage

\renewcommand{\theequation}{S-\arabic{equation}}
\renewcommand{\thefigure}{S\arabic{figure}}
\setcounter{equation}{0}
\setcounter{figure}{0}
\begin{widetext}
\begin{center}
{\Large \bf Supplemental material for: Phase transitions in definite total spin states of two-component Fermi gases}

Vladimir A. Yurovsky

\end{center}
\end{widetext}

\tableofcontents

\bigskip 

Numbers of equations and figures in the Supplemental material are started from S. References to equations and figures in the Letter do not contain S.

\section{Energy shifts in a cavity}
\begin{figure}[h]
\includegraphics[width=3.4in]{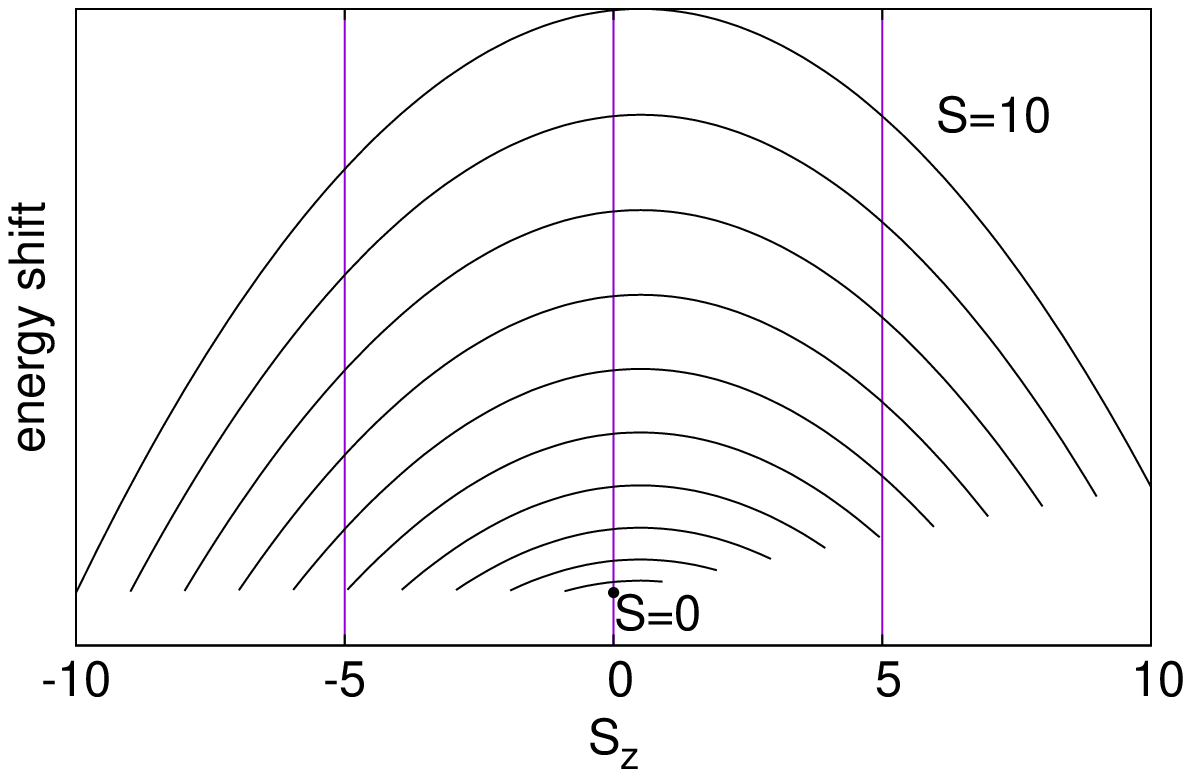}
 \caption{Energy shifts of states with different total spins $S$ due to an exchange of virtual photons in a two-dimensional cavity.   \label{Fig_cav_lev}}
\end{figure}

Spin-dependent spatially-homogeneous interactions of the form \cite{sela2016} 
$\hat{H}_{\mathrm{spin}}=I \hat{S}_+  \hat{S}_-$ appear in the collective Dicke states \cite{dicke1954} of two-level particles, coupled by electromagnetic field in a two-dimensional cavity. Such interaction  leads to the energy shift $I[S_z(S_z-1)-S(S+1)]$, providing substantial splitting of the states with different total spins (see Fig. \ref{Fig_cav_lev}). If $I>0$, the ground state of the system with given $S_z$ will be the state with the minimal allowed spin $S=S_z$ since $S$ cannot be less than $S_z$.

\section{States with well-defined many-body spins}
The spatial $\tilde{\Phi}_{t\hat{r}\{\varepsilon\}}^{(S)}$ and spin $\Xi_{t}^{(S)}$ wavefunctions in the eigenstate  \refeq{tilPsi} form the bases of irreducible representations of the symmetric group $\pr S_{N}$
of $N$-symbol permutations \cite{kaplan,pauncz_symmetric}. A permutation $\pr P$ of the particles transforms
each function to a linear combination of functions in the same representation,
\begin{equation}
\pr P\tilde{\Phi}_{t\hat{r}\{\varepsilon\}}^{(S)}  =  \sum_{t'}D_{t't}^{[\lambda]}(\pr P)
\tilde{\Phi}_{t'\hat{r}\{\varepsilon\}}^{(S)},
\quad
\pr P\Xi_{t}^{(S)}  =  \sum_{\tilde{t}'}D_{\tilde{t}'\tilde{t}}^{[\tilde{\lambda}]}(\pr P)\Xi_{t'}^{(S)}
\end{equation}
Here $D_{t't}^{[\lambda]}(\pr P)$ are the matrices of the Young orthogonal representation \cite{kaplan,pauncz_symmetric} associated with the Young diagrams $\lambda$. For spin-$1/2$ fermions,
the diagrams $\lambda=[2^{N/2-S}, 1^{2S}]$ have two columns and are unambiguously related to the total spin $S$. 
The representations of the spin and spatial wavefunctions are conjugate, and the dual Young diagrams $\tilde{\lambda}=[N/2+S,N/2-S]$ have two rows. The matrices of conjugate representations are related as $D_{\tilde{t}'\tilde{t}}^{[\tilde{\lambda}]}(\pr P)=\mathrm{sgn}(\pr P)D_{t't}^{[\lambda]}(\pr P)$, where $\mathrm{sgn}(\pr P)$ is the permutation parity, providing the proper permutation symmetry of the total wavefunction
$
\pr P\tilde{\Psi}_{\hat{r}\{\varepsilon\}}^{(S)}=\mathrm{sgn}(\pr P)\tilde{\Psi}_{\hat{r}\{\varepsilon\}}^{(S)}
$.
The representation functions are labeled by standard Young tableaux $t$ of the shape $\lambda$, the dual tableaux $\tilde{t}$ are obtained by replacing the rows with the columns. The representations have the dimension 
\begin{equation}
f_{S}(N)=\frac{N!(2S+1)}{(N/2+S+1)!(N/2-S)!}.
\label{fSN}
\end{equation}

The non-normalized spatial wavefunctions of non-interacting particles are expressed as
\begin{equation}
\tilde{\Phi}_{t\hat{r}\{\varepsilon\}}^{(S)}=\sum_{\pr P}D_{tr}^{[\lambda]}(\pr P)\prod_{j=1}^{N}\varphi_{\varepsilon_{j}}(\mathbf{r}_{\pr Pj}),
\label{tilPhi}
\end{equation}
where the relation between the Weyl tableau $\hat{r}$ and Young tableau $r$ is described below.
Single-body eigenfunctions $\varphi_{\varepsilon}(\mathbf{r})$  are solutions of the Schr\"odinger equation
\begin{equation}
\left[ \frac{\hbar^2}{2m}\hat{\mathbf{p}}_j^2+U(\mathbf{r}_j)\right] \varphi_{\varepsilon}(\mathbf{r})
=\varepsilon \varphi_{\varepsilon}(\mathbf{r})
\end{equation}
Here $\hat{\mathbf{p}}_j$ and $\mathbf{r}_j$ are momenta and coordinates of the fermions with the mass $m$ and $U(\mathbf{r}_j)$ is a spin-independent external potential. 

\section{Statistical weights of many-body states}
\begin{figure}
\begin{minipage}[t][][t]{1.5 in}
\vspace{0pt}
\includegraphics[width=1in]{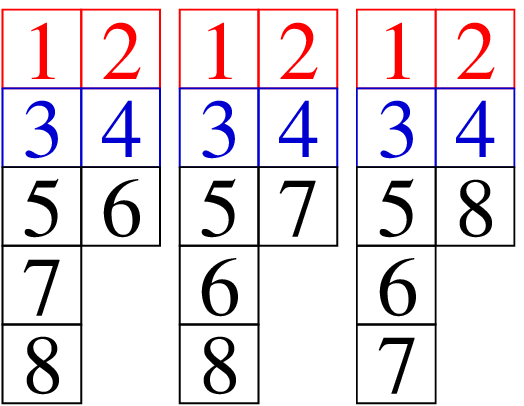} (a)
\end{minipage}
\begin{minipage}[t][][t]{1.5 in}
\vspace{0pt}
\includegraphics[width=1in]{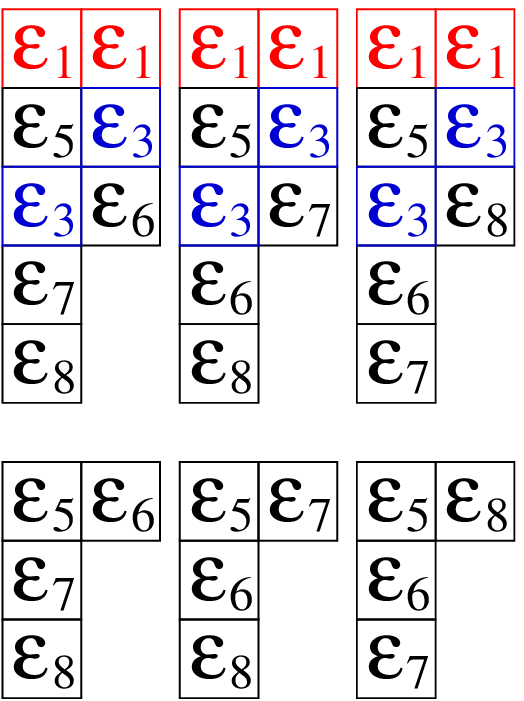} (b)
\end{minipage} 

\caption{(a) Three allowed standard Young tableaux for $N=8$, $S=1$, $q_2=2$, and $q_1=4$. The colored symbols correspond to double-occupied states and the black symbols, corresponding to single-occupied ones, fill the last rows, forming Young tableaux of the shape $[2,1^2]$. (b) Three allowed Weyl tableaux for $N=8$, $S=1$, $q_2=2$, $q_1=4$, and  $\varepsilon_1=\varepsilon_2<\varepsilon_5<\varepsilon_3=\varepsilon_4<\varepsilon_6<\varepsilon_7<\varepsilon_8$, obtained from the Young tableaux in part (a). The colored and black symbols correspond to double-occupied and single-occupied states, respectively. The standard Young tableaux of the shape $[2,1^2]$ in the second row are obtained from the Weyl tableaux by removal of the colored boxes. \label{Fig_Young_Weyl_nc}}
\end{figure}

According to \refeq{tilPhi}, each set of single-body states $\{\varepsilon\}$ provides several irreducible representations labeled by the standard Young tableaux $r$. A two-column Young diagram allows only single and double occupations of single-body states \cite{kaplan,yurovsky2014}. Let us suppose that the set $\{\varepsilon\}$  contains $q_2$ pairs ($\varepsilon_{2j}=\varepsilon_{2j-1}$ for $1\le j\le q_2$), corresponding to double occupied states, and $q_1=N-2q_2$ single-occupied states (it is clear, that physical consequences cannot depend on the state ordering). As demonstrated in \cite{yurovsky2014}, the first $q_2$ rows of $r$ have to contain two boxes each (this requires $q_2\leq N/2-S$) and have to be filled by the symbols $1\ldots 2q_2$. The $q_1$ symbols $2q_2+1\ldots N$ can occupy the remaining $N/2+S-q_2$ rows [see Fig. \ref{Fig_Young_Weyl_nc}(a)]. Then these rows form a  standard Young tableau of the shape $[2^{N/2-S-q_2},1^{2S}]$ and the number of such tableaux $f_S(q_1)$ is equal to the number of irreducible representations for the given set of single-body states.

Each of the tableaux $r$ can be unambiguously related to the Weyl (or semi-standard Young) tableau $\hat{r}$. The latter (see \cite{harshman2016a}) is a Young diagram filled by symbols such that they must increase down each column, but may remain the same or increase to the right in each row. The Weyl tableau $\hat{r}$ is obtained in the following way [see Fig. \ref{Fig_Young_Weyl_nc}(b)]: let us replace   $j$ by $\varepsilon_{j}$ in each box of the Young tableau $r$ and sort the entries in each column in the increasing down order  ($\varepsilon_{j'}$ can be less than $\varepsilon_{j}$ for $j'>j$ for the set $\{\varepsilon\}$ described above). 

Removing the boxes containing the degenerate energies $\varepsilon_{j}$ [see Fig. \ref{Fig_Young_Weyl_nc}(b)] one gets a standard Young tableau of the shape $[2^{N/2-S-q_2},1^{2S}]$. Then the number of the Weyl tableaux $\hat{r}$ (the Kostka number, see \cite{harshman2016a}) is equal to $f_S(q_1)$.

\section{Calculation of thermodynamic parameters}
The present adaptation of the Gentile's approach \cite{gentile1940} takes into account non-Abelian entropy. The most-probable values of the numbers $q^{(i)}_l$ of $l$-occupied levels ($0\leq l\leq 2$) in the $i$th cell are determined in the method of Lagrange undetermined multipliers by equations
\begin{multline}
\frac{\partial}{\partial q^{(i)}_l}\Bigl[  H + \alpha (N-\sum_i N_i)
+ \beta (E-\sum_i \bar{\varepsilon}_i N_i)
\\
+\sum_i  b_i \left( g_i-\sum_{l=0}^2 q_l^{(i)}\right) 
-\gamma'(q_2-N/2+S)\Bigr]=0 ,
\label{ObjFunc}
\end{multline}
where the entropy $H$ is given by \refeq{entropy}.  The consistency of the microstates with the macrostate,  expressed by conditions
\begin{equation}
\sum_i N_i=N,\quad \sum_i \bar{\varepsilon}_i N_i=E ,
\label{constraints}
\end{equation}
is provided by the Lagrange multipliers $\beta$ and $\alpha$. Here
\begin{equation}
N_i=q^{(i)}_1+2q^{(i)}_2 
\label{Ni}
\end{equation}
is the number of particles in the $i$th cell and $\bar{\varepsilon}_i$ is the average cell energy (see Fig. 1).The multipliers $b_i$ constrain the total number of levels in the $i$th cell
\begin{equation}
 g_i=\sum_{l=0}^2 q_l^{(i)}.
 \label{gi}
\end{equation} 
The multiplier $\gamma'$ is related to the inequality constraint (see \cite{nocedal}) on the total number of double-occupied levels $q_2=\sum_{i}q^{(i)}_2\leq N/2-S$. If $\gamma'>0$, the constraint is active and $q_2= N/2-S$ in the most-probable point. Otherwise, the constraint is inactive,  $\gamma'=0$, and the maximum of entropy, subject to constraints \refneq{constraints}, is attained at $q_2< N/2-S$. The transition between active and inactive inequality constraint is a mathematical description of the transition between saturated and unsaturated phases considered here. 

\begin{figure}[h]
\includegraphics[width=3.4in]{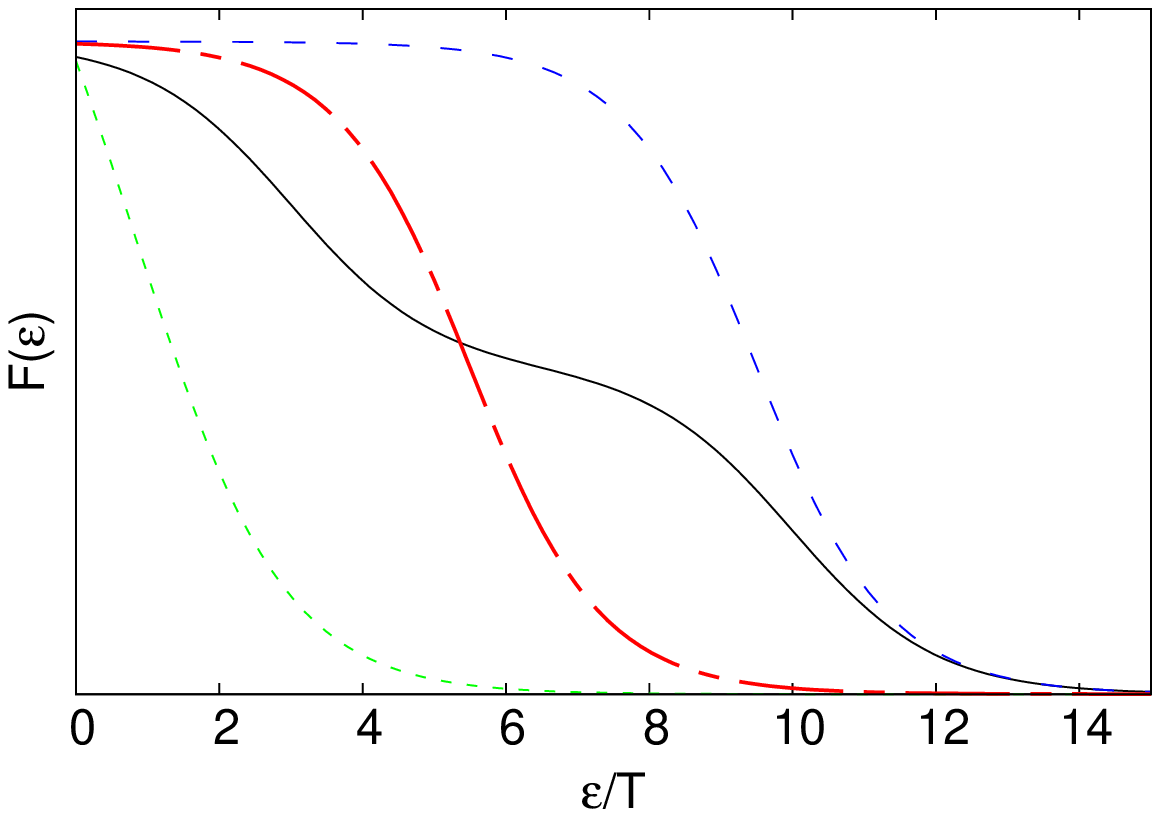}
 \caption{Energy distribution \refneq{Feps} for $\mu/T=10$ and $\gamma=7$ (black solid line), $\mu/T=10$ and $\gamma=1$ (blue long-dashed line), $\mu/T=1$ and $\gamma=4$ (green short-dashed line), and $\mu/T=6$ and $\gamma=1$ (red dot-dashed line). \label{Fig_fdistr}}
\end{figure}

Equations \refneq{ObjFunc} provide the numbers of $l$-occupied levels ($0\leq l\leq 2$) in the $i$th cell
\begin{equation}
 q^{(i)}_l=\exp(-\alpha l -\beta \bar{\varepsilon}_i l - \gamma \delta_{l 2} -b_i -1),
 \label{giexp}
\end{equation} 
where
\begin{equation}
 \gamma=\gamma'+2\frac{\partial \ln f_S(q_1)}{\partial q_1}.
 \label{gamma}
\end{equation}
Substitution of \refeq{giexp} into \refeq{gi} determines the  Lagrange multipliers $b_i$. Then using \refeq{Ni}, one gets
the energy-distribution function $F(\bar{\varepsilon}_i)=N_i/g_i$ 
\begin{equation}
 F(\varepsilon)=\frac{e^{-(\varepsilon-\mu)/T}
 +2e^{-2(\varepsilon-\mu)/T-\gamma}}
 {1+e^{-(\varepsilon-\mu)/T}+e^{-2(\varepsilon-\mu)/T-\gamma}}.
 \label{Feps}
\end{equation}
Here the Lagrange multipliers $\beta=1/T$ and $\alpha=-\mu/N$ are, usually, related to the temperature $T$ and the chemical potential $\mu$. The distribution depends on an additional parameter $\gamma$ [see \refeq{gamma}].
Examples of the energy distributions are presented in Fig. \ref{Fig_fdistr}. 

\newpage
The number of particles and the energy are calculated in the approximation of the continuous energy spectrum as
\begin{equation}
N(\mu,\gamma)=\int_0^\infty d \varepsilon \nu(\varepsilon) F(\varepsilon),
\quad
E(\mu,\gamma)=\int_0^\infty d \varepsilon \nu(\varepsilon) \varepsilon F(\varepsilon)
\label{N_E_mugam}
\end{equation}
Here $\nu(\varepsilon)$ is the energy-density of single-body levels. The present work deals with $\nu(\varepsilon)=\nu_k \varepsilon^k$, where $\nu_k$ are taken from \cite{pitaevskii,pethick}. In the case of a flat potential, $k=0$, $\nu_0=m V_{2D}/(2\pi\hbar^2)$ for a two-dimensional (2D) gas constrained in the area (two-dimensional volume) $V_{2D}$ and $k=1/2$, $\nu_{1/2}=m^{3/2}V_{3D}/(\sqrt{2}\pi^2\hbar^3)$ for a three-dimensional (3D) gas constrained in the volume $V_{3D}$. For anisotropic harmonic trapping, we have $k=1$, $\nu_1=(\hbar\omega)^{-2}$ and $k=2$, $\nu_2=(\hbar\omega)^{-3}/2$ in the 2D and 3D cases, respectively, where $\omega$ is the average angular frequency of the trap. 
An isolated 1D gas in a harmonic axial potential does not demonstrate thermalization \cite{kinoshita2006} and can be described by a thermodynamic ensemble only due to interactions with the environment. This system corresponds to $k=0$ with $\nu_0=1/(\hbar\omega)$. 
The derivative of the non-Abelian entropy
\begin{multline}
 \tilde{\gamma}_S(q_1)=2\frac{\partial \ln f_S(q_1)}{\partial q_1}
 \\
 =2\psi(q_1+1)-\psi(\frac{q_1}{2}+S+2)-\psi(\frac{q_1}{2}-S+1)
\end{multline}
is expressed by differentiation of \refeq{fSN} in terms of the logarithmic derivative of the $\Gamma$ function $\psi$ (see \cite{olver}). 
In the approximation of the continuous energy spectrum, the number of single-occupied levels is expressed as
\begin{equation}
q_1(\mu,\gamma)=\int_0^\infty d \varepsilon \nu(\varepsilon)
\frac{e^{-(\varepsilon-\mu)/T}}
 {1+e^{-(\varepsilon-\mu)/T}+e^{-2(\varepsilon-\mu)/T-\gamma}} .
 \label{q1mugam}
\end{equation}
See Sec. \ref{SI_integrals} for details of the integral calculation in Eqs. \refneq{N_E_mugam} and \refneq{q1mugam}.

\begin{figure}
\includegraphics[width=3.4in]{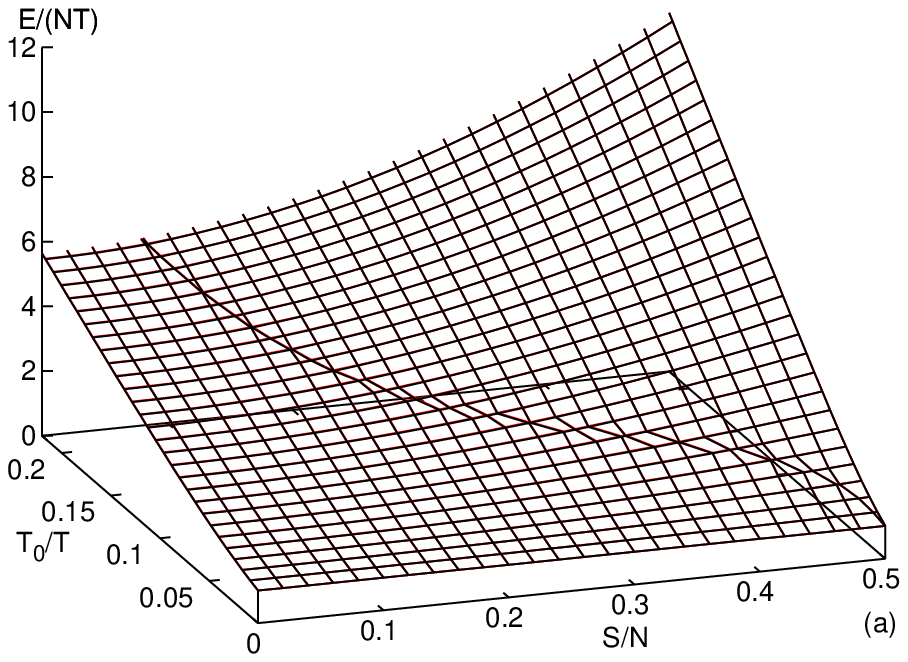}
\includegraphics[width=3.4in]{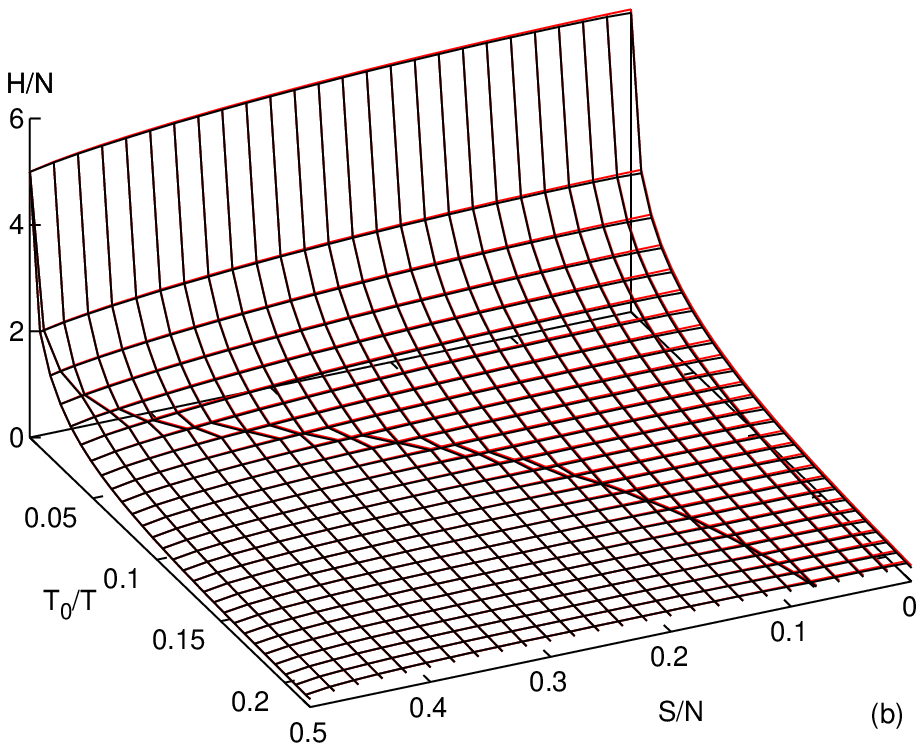}
 \caption{The energy (a) and the entropy (b) at the temperature $T$ for the state with the defined many-body spin $S$ (black) and the state with defined individual spins and the total spin projection $S$ (red) of $N=10^2$ two-dimensional particles in flat potential. The red and black plots are almost indistinguishable. \label{Fig_E_H}}
\end{figure}

\clearpage
Given $N$, $S$, and $T$, the parameters $\mu$ and $\gamma$ are solutions of the equations
\begin{equation}
N(\mu,\gamma)=N,
\quad q_1(\mu,\gamma)=2S
\label{mugamconstr}
\end{equation}
[see \refeq{N_E_mugam}]
in the saturated phase, when $\gamma>\tilde{\gamma}_S(q_1(\mu,\gamma))$, i. e. $\gamma'>0$. Otherwise, the gas is in the unsaturated phase, $\gamma'=0$, and $\mu$ and $\gamma$ are solutions of the equations
\begin{equation}
N(\mu,\gamma)=N,
\quad \tilde{\gamma}_S(q_1(\mu,\gamma))=\gamma .
\label{mugamnoncon}
\end{equation}
Having $\mu$ and $\gamma$ we can calculate the energy 
with \refeq{N_E_mugam}. Then \refeq{entropy} gives us the entropy
\begin{equation}
H=\frac{k+2}{k+1}\frac{E}{T}-\frac{\mu N}{T}+\gamma\frac{N-q_1}{2}+\ln f_S(q_1)
\end{equation}
The last two terms here are related to the non-Abelian permutation symmetry.
The energy and entropy are continuous (see Fig. \ref{Fig_E_H}).

The pressure is calculated as a derivative of the energy over volume with fixed entropy, taking into account that $\partial \nu_k/\partial V=\nu_k/V$ for flat potentials, where volume is defined. Then
\begin{equation}
 P=-\left( \frac{\partial E}{\partial V}\right)_H=-\left( \frac{\partial E}{\partial V}\right)_T+T \left( \frac{\partial H}{\partial V}\right)_T=\frac{1}{k+1}\frac{E}{V},
\end{equation} 
in agreement with the general relations for ideal gases, since the non-Abelian contributions in derivatives of energy and entropy are canceled.

\begin{figure}[h]
\includegraphics[width=3.4in]{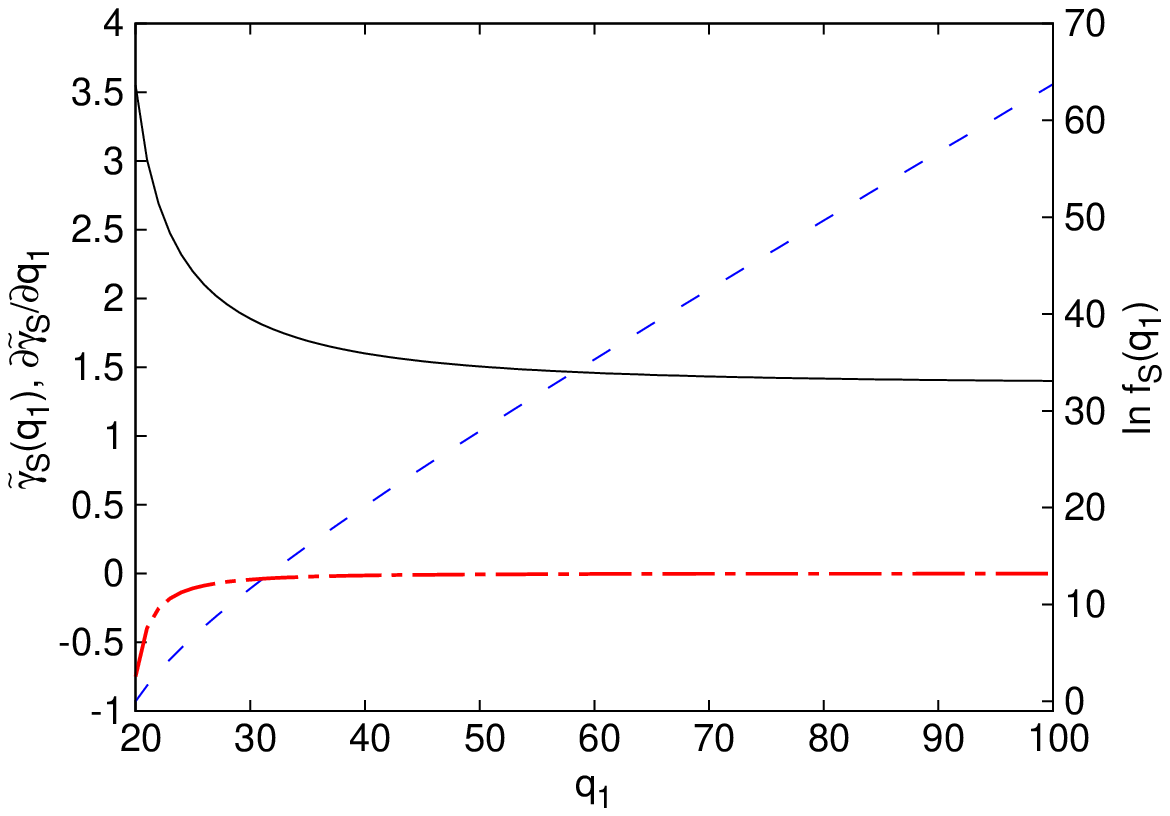}
 \caption{Dependencies of $\ln f_S(q_1)$ (blue dashed line), $\tilde{\gamma}_S(q_1)$ (black solid line), and $\frac{\partial \tilde{\gamma}_S(q_1)}{\partial q_1}$ (red dot-dashed line) on $q_1$ for $S=10$.   \label{Fig_fS_der}}
\end{figure}

\begin{figure}[!]
\includegraphics[width=3.4in]{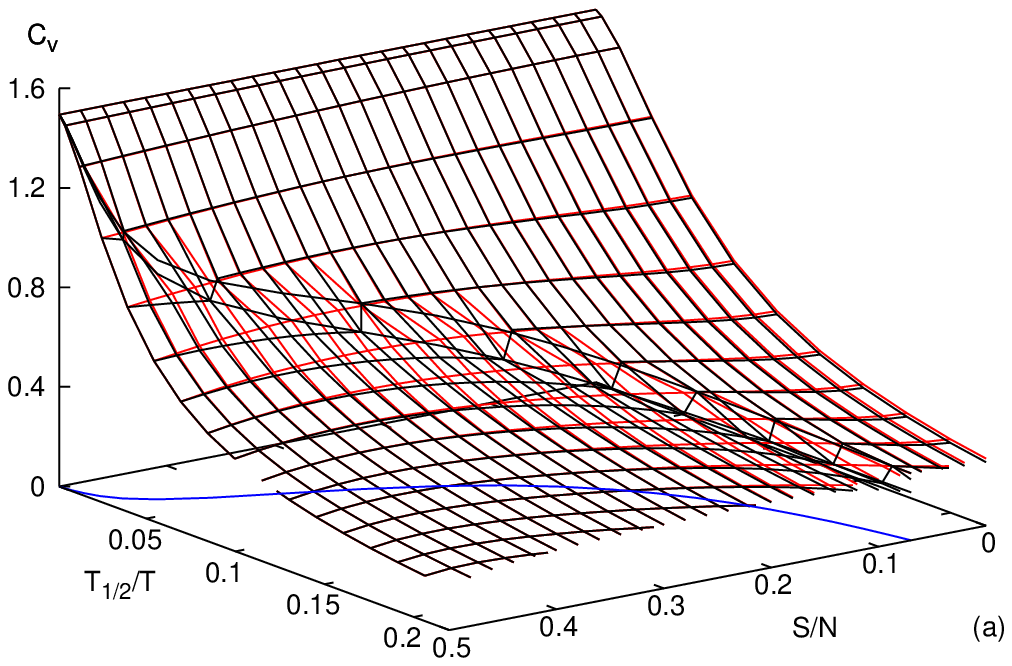}
\includegraphics[width=3.4in]{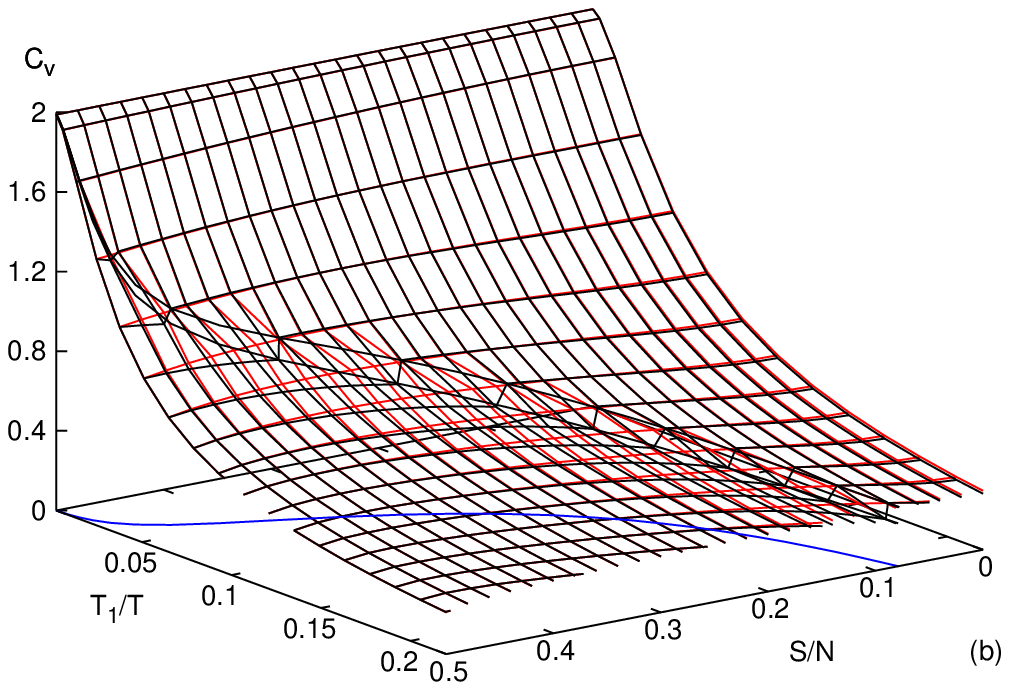}
\includegraphics[width=3.4in]{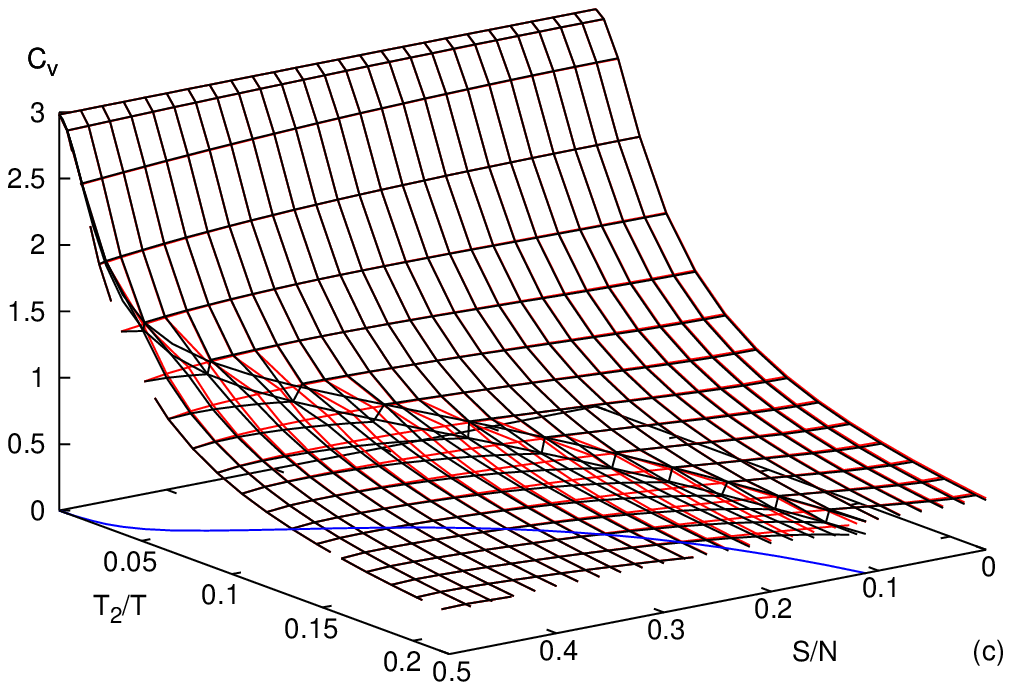}
 \caption{Specific heat (per atom) at the temperature $T$ for the state with the defined many-body spin $S$ (black) and the state with defined individual spins and a total spin projection $S$ (red) of $N=10^2$ particles. The blue line shows the boundary between the phases. (a) For a three-dimensional gas in a flat potential. (b) For a two-dimensional harmonic confinement. (c) For a three-dimensional harmonic confinement. \label{Fig_f3_h2_h3}}
\end{figure}

\clearpage
The specific heat (per atom) $C_V$ is defined for flat potentials as a derivative of the energy over the temperature at the constant volume divided by $N$. For trapped gases, it is defined as the derivative for the fixed trap potential, as in  \cite{pethick}.
The specific heat is expressed as
\begin{equation}
C_V=\frac{1}{N}\left( \frac{\partial E}{\partial T}
 +\frac{\partial E}{\partial \mu} \frac{\partial \mu}{\partial T}
 +\frac{\partial E}{\partial \gamma} \frac{\partial \gamma}{\partial T}\right) .
\end{equation}
In the saturated phase, equations for the derivatives of $\mu$ and $\gamma$
\begin{equation}
 \frac{\partial N}{\partial \mu}\frac{\partial \mu}{\partial T}
 +\frac{\partial N}{\partial \gamma}\frac{\partial \gamma}{\partial T}=-\frac{\partial N}{\partial T},
\quad
\frac{\partial q_1}{\partial \mu}\frac{\partial \mu}{\partial T}
 +\frac{\partial q_1}{\partial \gamma}\frac{\partial \gamma}{\partial T}=-\frac{\partial q_1}{\partial T}
\end{equation}
are obtained by differentiation of \refeq{mugamconstr}. In the unsaturated phase, due to \refeq{mugamnoncon}, the second equation is modified as
\begin{equation}
\frac{\partial q_1}{\partial \mu}\frac{\partial \mu}{\partial T}
 +\left[ \frac{\partial q_1}{\partial \gamma}
- \left( \frac{\partial \tilde{\gamma}_S(q_1)}{\partial q_1}\right)^{-1} \right] 
\frac{\partial \gamma}{\partial T}=-\frac{\partial q_1}{\partial T}.
\end{equation}
Here
\begin{equation}
 \frac{\partial \tilde{\gamma}_S(q_1)}{\partial q_1}=
 2\psi'(q_1+1)-\frac{1}{2}\psi'(\frac{q_1}{2}+S+2)-\frac{1}{2}\psi'(\frac{q_1}{2}-S+1).
\end{equation}
and $\psi'$ is the trigamma function (see  \cite{olver}). Since $\partial \tilde{\gamma}_S(q_1=2S)/\partial q_1\neq 0$ (see Fig. \ref{Fig_fS_der}), the derivatives of $\mu$ and $\gamma$ have discontinuities at $q_1=2S$. This leads to discontinuity of $C_V$ shown in Figs. 2, 3 and \ref{Fig_f3_h2_h3}. 

\section{Expectation values for thermalized eigenstates}\label{SI_EigenTherm}
The alternative derivation, presented below, is not based on the postulate of equal \textit{a priory} probabilities. It is applicable to gases in flat potentials of any dimension. In the chaotic regime, according to the Berry conjecture  \cite{berry1977}, each eigenstate appears to be a superposition of plane waves with random phases and Gaussian random amplitudes, but with fixed energies. In the Srednicki form  \cite{srednicki1994}, the spatial wavefunction of interacting particles is expressed as 
\begin{multline}
\Phi_{tn}^{(S)}=\mathcal{N}_{n}^{(S)}\hat{W}_{tn}^{(S)}\sum_{\{\mathbf{p}\}}A_{n}(\{\mathbf{p}\})\tilde{\delta}(\{\mathbf{p}\}^{2}-2mE_{n}^{(S)})
\\
\times
\exp(i\sum_{j}\mathbf{p}_{j}\mathbf{r}_{j}/\hbar),
\label{PhiStn}
\end{multline}
where $\{\mathbf{p}\}\equiv\{\mathbf{p}_{1},\ldots,\mathbf{p}_{N}\}$
 is the set of particle momenta in the periodic box of the volume $V$ with incommensurable dimensions and $\{\mathbf{p}\}^{2}\equiv\sum_{j}\mathbf{p}_{j}^{2}$. Since the momenta $\mathbf{p}_{j}$ have a discrete spectrum, the states with approximately fixed energies $E_{n}^{(S)}$ are selected by the function
\[
\tilde{\delta}(x)=\Theta(\Delta-|x|)/(2\Delta),
\]
where $\Theta(x)$ is the Heaviside step function. The Gaussian
random coefficients $A_{n}(\{\mathbf{p}\})$ have the two-point correlation function 
\begin{equation}
\left\langle A_{n'}^{*}(\{\mathbf{p'}\})A_{n}(\{\mathbf{p}\})\right\rangle _{\mathrm{EE}}=\frac{\delta_{n'n}\delta_{\{\mathbf{p'}\}\{\mathbf{p}\}}}{\tilde{\delta}(\{\mathbf{p'}\}^{2}-\{\mathbf{p}\}^{2})},
\label{Acorr}
\end{equation}
where $\left\langle \right\rangle _{\mathrm{EE}}$
denotes the average over a fictitious ``eigenstate ensemble'', which
describes properties of a typical eigenfunction \cite{srednicki1994}.

Generalizing the Srednicki treatment \cite{srednicki1994} of symmetric and anti-symmetric wavefunctions to non-Abelian representations, the proper permutation symmetry is provided by the symmetrization operator
\begin{equation}
\hat{W}_{tn}^{(S)}=\sum_r B_{nr}^{(S)}\sum_{\pr P}D_{tr}^{[\lambda]}(\pr P)\pr P.
\end{equation}
Any choice of the factors $B_{nr}^{(S)}$ leads to 
$\pr P\Phi_{tn}^{(S)}  =  \sum_{t'}D_{t't}^{[\lambda]}(\pr P)\Phi_{t'n}^{(S)}$. Then the total wavefunction $\Psi_{n}^{(S)}=f_{S}^{-1/2}(N)\sum_{t}\Phi_{tn}^{(S)}\Xi_{t}^{(S)}$ has the proper fermionic permutation symmetry. Without loss of generality, we can suppose $\sum_r|B_{nr}^{(S)}|^2=1$.

Unlike \cite{yurovsky2016}, the wavefunction \refneq{PhiStn} does not neglect multiple occupations of the momentum states. It can be represented as
\begin{multline}
\Phi_{tn}^{(S)}=\mathcal{N}_{n}^{(S)}\sum_{\{\mathbf{p}\}}A_{n}^{(S)}(t,\{\mathbf{p}\})
\tilde{\delta}(\{\mathbf{p}\}^{2}-2mE_{n}^{(S)})
\\
\times
\exp(i\sum_{j}\mathbf{p}_{j}\mathbf{r}_{j}/\hbar)
\end{multline}
with
\begin{equation}
A_{n}^{(S)}(t,\{\mathbf{p}\})=\sum_r B_{nr}^{(S)}\sum_{\pr P}D_{tr}^{[\lambda]}(\pr P)A_{n}(\{\mathbf{p}_{\pr P j}\}).
\end{equation}
Due to orthogonality of the spin wavefunctions, the expectation values of a symmetric one-body spin-independent operator $\sum_j \hat{O}(\mathbf{r}_{j})$ in the eigenstate $\Psi_{n}^{(S)}$ is reduced to the expectation values in the spatial states,
\begin{equation}
\langle \Psi_{n}^{(S)}|\sum_j \hat{O}(\mathbf{r}_{j})|\Psi_{n}^{(S)}\rangle=\frac{1}{f_S(N)}
\sum_{t}\langle \Phi_{tn}^{(S)}|\sum_j \hat{O}(\mathbf{r}_{j})|\Phi_{tn}^{(S)}\rangle ,
\end{equation}
which can be estimated by their eigenstate-ensemble averages,
\begin{multline}
\left\langle \langle \Phi_{t'n}^{(S)}|\sum_j \hat{O}(\mathbf{r}_{j})|\Phi_{tn}^{(S)}\rangle\right\rangle _{\mathrm{EE}}=\left( \mathcal{N}_{n}^{(S)}\right)^2 V^{N-1}
\\
\times
\sum_{\{\mathbf{p}\},\{\mathbf{p}'\}}\sum_j 
\langle \exp(i\mathbf{p}'_{j}\mathbf{r}_{j}/\hbar)|\hat{O}(\mathbf{r}_{j})|\exp(i\mathbf{p}_{j}\mathbf{r}_{j}/\hbar)\rangle 
\\
\times
\prod_{j'\neq j}\delta_{\mathbf{p}'_{j'}\mathbf{p}_{j'}}
\tilde{\delta}(\{\mathbf{p}\}^{2}-2mE_{n}^{(S)})
\sum_{r,r'}\left(B_{nr'}^{(S)}\right)^*B_{nr}^{(S)}
\\
\times
\sum_{\pr P,\pr P'}D_{t'r'}^{[\lambda]}(\pr P')D_{tr}^{[\lambda]}(\pr P)
\prod_{j''} \delta_{\mathbf{p'}_{\pr P' j''}\mathbf{p}_{\pr P j''}}
\end{multline}
The last product of the Kronecker symbols, originated from the correlation function \refneq{Acorr}, means that each element of  the set $\{\mathbf{p}'\}$ is equal to any element of the set $\{\mathbf{p}\}$. Moreover, the first product, which originate from the orthogonality of the plane waves, means that all but one $\mathbf{p}'_j$ are equal to $\mathbf{p}_j$. Therefore, $\mathbf{p}'_j=\mathbf{p}_j$ for any $j$ and $\pr P'=\pr P^{\{N\}}(\{\mathbf{p}\})\pr P$, where the permutations 
$\pr P^{\{N\}}(\{\mathbf{p}\})$ (cf \cite{yurovsky2014}) do not affect the set $\{\mathbf{p}\}$, permuting only the equal elements, $\mathbf{p}_{\pr P^{\{N\}}j}=\mathbf{p}_j$. Then the sum over $\pr P$ and $\pr P'$ can be transformed as
\begin{multline}
\sum_{\pr P^{\{N\}}}\sum_{t''}D_{t't''}^{[\lambda]}(\pr P^{\{N\}})
\sum_{\pr P}D_{t''r'}^{[\lambda]}(\pr P)D_{tr}^{[\lambda]}(\pr P)
\\
=\frac{N!}{f_S(N)}\delta_{rr'}\sum_{\pr P^{\{N\}}}D_{t't}^{[\lambda]}(\pr P^{\{N\}}).
\end{multline}
Here the general relation for representation matrices \cite{kaplan,pauncz_symmetric}
\begin{equation}
\sum_{t}D_{r't}^{[\lambda]}(\pr P)D_{tr}^{[\lambda]}(\pr Q)=D_{r'r}^{[\lambda]}(\pr P\pr Q)
\label{RepresMat}
\end{equation}
and the orthogonality relation
\begin{equation}
\sum_{\pr P}D_{t'r'}^{[\lambda']}(\pr P)D_{tr}^{[\lambda]}(\pr P)=
\frac{N!}{f_{S}(N)}\delta_{tt'}\delta_{rr'}\delta_{\lambda\lambda'}
\end{equation}
are used.

Finally, we get
\begin{multline}
\left\langle \langle \Psi_{n}^{(S)}|\sum_j \hat{O}(\mathbf{r}_{j})|\Psi_{n}^{(S)}\rangle\right\rangle _{\mathrm{EE}}=\frac{N!}{f_S^2(N)}\left( \mathcal{N}_{n}^{(S)}\right)^2 V^{N}
\\
\times
\sum_{\{\mathbf{p}\}}
\tilde{\delta}(\{\mathbf{p}\}^{2}-2mE_{n}^{(S)})\sum_j\bar{O}(\mathbf{p}_{j})
\sum_{t,\pr P^{\{N\}}}D_{tt}^{[\lambda]}(\pr P^{\{N\}}(\{\mathbf{p}\})),
\label{PsiOPsi1}
\end{multline}
where
\begin{equation}
\bar{O}(\mathbf{p}_{j})=\frac{1}{V}\langle \exp(i\mathbf{p}_{j}\mathbf{r}_{j}/\hbar)|\hat{O}(\mathbf{r}_{j})|\exp(i\mathbf{p}_{j}\mathbf{r}_{j}/\hbar)\rangle 
\end{equation}

A state associated with the two-column Young diagram $r$ cannot have more than two equal momenta \cite{kaplan,yurovsky2014}. Let $\mathbf{p}_{i'_k}=\mathbf{p}_{i''_k}$ for $1\leq k\leq q_2$. Then 
$\pr P^{\{N\}}(\{\mathbf{p}\})$ can be either the identity permutation $\pr E$ or any product of transpositions 
$\pr P_{i'_ki''_k}$. It can be represented as 
$\pr P^{\{N\}}(\{\mathbf{p}\})=\pr Q\tilde{\pr P}^{\{N\}}\pr Q^{-1}$, where $\pr Q(2k-1)=i'_k$, $\pr Q(2k)=i''_k$, and $\tilde{\pr P}^{\{N\}}$ can be either $\pr E$ or any product of transpositions $\pr P_{2k-1,2k}$. Then \refeq{RepresMat} allows us to transform the sum of Young orthogonal matrices in \refeq{PsiOPsi1} in the following way
\begin{multline}
\sum_{t,t'}D_{tt'}^{[\lambda]}(\pr Q)D_{t't}^{[\lambda]}(\tilde{\pr P}^{\{N\}}\pr Q^{-1})=
\sum_{t'}D_{t't'}^{[\lambda]}(\tilde{\pr P}^{\{N\}}\pr Q^{-1}\pr Q)
\\
=\sum_{t}D_{tt}^{[\lambda]}(\tilde{\pr P}^{\{N\}}).
\end{multline}
As demonstrated in \cite{yurovsky2014}, $\sum_{\tilde{\pr P}^{\{N\}}}D_{tt}^{[\lambda]}(\tilde{\pr P}^{\{N\}})$ vanishes unless the first $2q_2$ symbols occupy first $q_2$ rows in the Young tableau $t$. Thus only the last $N/2-S-q_2$ rows of this tableau can be changing in the sum over $t$. These rows form a standard Young tableau of the shape $[2^{N/2-S-q_2},1^{2S}]$, filled by the symbols $2q_2+1\ldots N$. Since permutations $\tilde{\pr P}^{\{N\}}$ do not affect these symbols, $D_{tt}^{[\lambda]}(\tilde{\pr P}^{\{N\}})=1$ and $\sum_{t}D_{tt}^{[\lambda]}(\tilde{\pr P}^{\{N\}})=f_S(q_1)$. There are $2^{q_2}$ permutations $\tilde{\pr P}^{\{N\}}$. Therefore,
\begin{multline}
\left\langle \langle \Psi_{n}^{(S)}|\sum_j \hat{O}(\mathbf{r}_{j})|\Psi_{n}^{(S)}\rangle\right\rangle _{\mathrm{EE}}=\frac{N!}{f_S^2(N)}\left( \mathcal{N}_{n}^{(S)}\right)^2 V^{N}
\\
\times
\sum_{\{\mathbf{p}\}}2^{q_2}f_S(q_1)\tilde{\delta}(\{\mathbf{p}\}^{2}-2mE_{n}^{(S)})
\sum_j\bar{O}(\mathbf{p}_{j}) .
\end{multline}

Let us divide the single-body energy-spectrum into cells, as it was done on derivation  of \refeq{entropy}, and suppose that $\bar{O}(\mathbf{p}_{j})$ can be approximated by $\bar{O}_i$ in the $i$ th energy cell. Then the summation over $\{\mathbf{p}\}$ can be replaced by summation over numbers of non-, single-, and double-occupied levels ($q^{(i)}_0$, $q^{(i)}_1$, and $q^{(i)}_2$, respectively) in each cell. The levels can be distributed in $\prod_ig_i!/(q^{(i)}_0!q^{(i)}_1!q^{(i)}_2!)$ distinct ways and particles can be distributed in $N!/2^{q_2}$ distinct ways between the occupied levels. Then
\begin{multline}
\left\langle \langle \Psi_{n}^{(S)}|\sum_j \hat{O}(\mathbf{r}_{j})|\Psi_{n}^{(S)}\rangle\right\rangle _{\mathrm{EE}}=\left( \frac{N!}{f_S(N)}\mathcal{N}_{n}^{(S)}\right)^2 V^{N}
\\
\times
\sum_{\{q^{(i)}_l\}}f_S(q_1)\sum_iN_i\bar{O}_i \prod_i\frac{g_i!}{q^{(i)}_0!q^{(i)}_1!q^{(i)}_2!}
\\
\times
\tilde{\delta}(E-\sum_i \bar{\varepsilon}_i N_i)\delta_{N,\sum_i N_i}\prod_i\delta_{g_i,\sum_lq_l^{(i)}}
\end{multline}
(recall, that $N_i=\sum_{l=1}^2 l q_l^{(i)}$). The factor $f_S(q_1)$, providing the non-Abelian entropy, appears here, although the states of interacting particles $\Psi_{n}^{(S)}$ have no defined set of single-body states and are not labeled by Weyl tableaux. The sum can be approximated by its dominant term, corresponding to the maximum of the entropy \refneq{entropy}. 

\section{Calculation of integrals}\label{SI_integrals}
Whenever $\gamma>\ln 4$, the integrals in Eqs. \refneq{N_E_mugam} and \refneq{q1mugam} for $E$, $N$, and $q_1$ can be expressed in terms of  the Fermi-Dirac function \cite{olver}
\[
 F_k(y_0)=\int_0^\infty \frac{y^k dy}{1+e^{y-y_0}}
\]
as
\[
 N=\nu_k T^{k+1}\left( F_k(y_1)+F_k(y_2)\right)
\]
\[
 E=\nu_k T^{k+2}\left( F_{k+1}(y_1)+F_{k+1}(y_2)\right)
\]
\[
 q_1=\nu_k T^{k+1}\left( 1-4e^{-\gamma}\right)^{-1/2}\left( F_k(y_1)-F_k(y_2)\right), 
\]
where
\[
 y_{1,2}=\frac{\mu}{T}+\ln\frac{1\pm\sqrt{1-4e^{-\gamma}}}{2}
\]
The Fermi-Dirac function is calculated with the code \cite{gong2001}. For $\gamma\leq\ln 4$, direct numerical integration is used in Eqs. \refneq{N_E_mugam} and \refneq{q1mugam} for $E$, $N$, and $q_1$.

In the case of two-dimensional gas in a flat potential, due to homogeneity of the single-body energy spectrum, some integrals can be calculated analytically. Substituting $\varepsilon=\mu-T\ln x$ we get from equations \refneq{Feps} and \refneq{N_E_mugam}
\begin{multline*}
 N(\mu,\gamma)=\nu_0 T\int_0^{\exp(\mu/T)} dx \frac{1+2xe^{-\gamma}}{1+x+x^2e^{-\gamma}}
 \\
 =\nu_0 T\ln\left( 1+e^{\mu/T}+e^{2\mu/T-\gamma}\right) .
\end{multline*}
Equation \refneq{q1mugam} can be transformed as
\begin{multline*}
 q_1(\mu,\gamma)=\nu_0 T\int_0^{\exp(\mu/T)}  \frac{dx}{1+x+x^2e^{-\gamma}}
 \\
  =\nu_0 T \begin{cases}
           \frac{2}{v} \arctan \frac{v}{1+2e^{-\mu/T}} &, \gamma<\ln 4 \\
           \frac{2}{1+2e^{-\mu/T}} &, \gamma=\ln 4 \\
           \frac{2}{|v|}\mathrm{arctanh}\frac{|v|}{1+2e^{-\mu/T}} &, \gamma>\ln 4 
           \end{cases}
\end{multline*}
where $v=\sqrt{4e^{-\gamma}-1}$.

\section{Applicability criteria}\label{SI_AppCrit}
Spin and spatial degrees of freedom become inseparable in the presence of spin-dependent two-body interactions. In the case of cold fermionic atoms, this can be caused by the inapplicability of the zero-range approximation. At the low atom energy $E$ the elastic scattering amplitude (see \cite{landau})
\begin{equation}
 f=\left[ -\frac{1}{a_S}-i \frac{\sqrt{m E}}{\hbar}+\frac{m E}{2\hbar^2}r_{\text{eff}}\right]^{-1} 
\end{equation} 
is expressed in terms of the $S$-wave elastic scattering length $a_S$ and the effective interaction radius $r_{\text{eff}}$. The correction to the zero-range approximation --- the term, proportional to $r_{\text{eff}}$ --- is negligible whenever $E\ll 4\hbar^2/(m r_{\text{eff}}^2)$. The applicability condition for the de Broglie wavelength $\lambda_F\equiv 2 \pi\hbar/\sqrt{2 m E}$ is  $\lambda_F\gg \pi r_{\text{eff}}/\sqrt{2}$.  

For the van der Waals interaction potential the scattering length and effective radius are expressed (see \cite{landau,chin2010}) in terms of the van der Waals radius $R_{\text{vdW}}$,
\begin{align*}
 a_S=&\frac{\Gamma(3/4)}{\Gamma(5/4)}R_{\text{vdW}}\approx 1.35 R_{\text{vdW}}
 \\
 r_{\text{eff}}=&\frac{2}{3\pi}\left[ 1-\frac{8\pi\Gamma(5/4)}{\Gamma^2(1/4)\Gamma(3/4)}
 +2\left( \frac{4\pi\Gamma(5/4)}{\Gamma^2(1/4)\Gamma(3/4) }\right)^2\right]
 \\
 &\times\Gamma^2(1/4)R_{\text{vdW}}
 \approx 1.63 R_{\text{vdW}} .
\end{align*}
Then the condition for the atom energy is expressed in terms of the van der Waals energy \cite{chin2010}
$E_{\text{vdW}}=\hbar^2/(m R_{\text{vdW}}^2)$ as $E\ll 1.5 E_{\text{vdW}}$. Using the values of 
$R_{\text{vdW}}\approx 31.26 a_B$ (where $a_B$ is the Bohr radius) and $E_{\text{vdW}}\approx 29.47$ mK \cite{chin2010} for ${}^6$Li, one gets the condition $E\ll 44$mK.

Thermodynamic ensemble predictions are applicable to open systems, thermalized due to interactions with the environment, as well as to chaotic isolated systems, when eigenstate thermalization takes place \cite{srednicki1994}. For isolated systems, the criterion of chaos in a gas with zero-range interactions is $\lambda_F<a_S N^2$ (see \cite{yurovsky2016}). In the temperature range of interest $T_k<T<N T_k$, the characteristic energy is $\varepsilon_F$, $\lambda_F=2 \pi\hbar/\sqrt{2 m\varepsilon_F}$, and both criteria can be expressed as $r_{\text{eff}}N^{1/3}\ll V_{3D}^{1/3}< |a_S| N^{7/3}$ for a 3D gas in a flat potential, constrained in the volume $V_{3D}$. The gases can have a 2D behavior under a strong axial confinement with the range $a_{\text{conf}}\ll \lambda_F$. Then the criteria are expressed in terms of $a_{\text{conf}}$ as $r_{\text{eff}}\ll a_{\text{conf}}<|a_S|N^2$ \cite{yurovsky2016}. Relative fluctuations for the 3D case can be estimated as in \cite{srednicki1994}, $N^{1/2}(\lambda_F V_{3D}^{-1/3})^{3N/2-3}\approx N^{3/2-N/2}$. For a 2D gas in a flat potential, a similar estimation gives $N^{5/4-N/2}$. The fluctuations are negligibly small for $N\gtrsim 10$. 

The present theory is applicable to weakly-interacting gases. In the method of cluster expansions \cite{pathria} and theory of quantum gases \cite{pitaevskii}, the effect of interactions is characterized by the parameters $a_S/\lambda_F$ and $a_S (N/V_{3D})^{1/3}$. Since $|a_S|\approx r_{\text{eff}}$ for non-resonant interactions, both parameters are small when the zero-range approximation is applicable.
Interactions can also lead to the formation of dimers or Cooper pairs for repulsive or attractive interactions, respectively. The present analysis neglects these effects, being applicable to so-called ``upper branch BEC'' for repulsive interactions, where the particles do not form bound states, or to non-superfluid regime for attractive interactions. This is justified by the superfluidity transition temperature $\exp(-\text{const}\lambda_F/|a_S|)$ (see \cite{pathria}), which is negligibly small. Under the same conditions for repulsive interactions, the dimer states can be neglected since their binding energy $\sim \hbar^2/(m a_S^2)$ substantially exceed $\varepsilon_F$.

Discontinuities in the thermodynamic functions exhibited by phase transitions can appear only in infinite systems (see \cite{pathria}). In finite systems, the thermodynamic functions are continuous, although sharp changes can appear. In the present case, the corner of the non-Abelian entropy $\ln f_S(q_1)$ at the critical temperature (see Fig. 2) is smoothed due to fluctuations of the number of single-occupied levels $q_1$. Then the discontinuity of the specific heat is transformed to a continuous crossover. However, the approach \cite{srednicki1994} applied to fluctuations of $q_1$ provides the upper limit of relative fluctuations which is proportional to $N^{-N/2}$. The crossover temperature range has the same $N$ dependence.  
\end{document}